\documentclass[aps,pra,twocolumn,showpacs,floatfix,preprintnumbers,amsmath,amssymb]{revtex4-2}

\usepackage[english]{babel}
\usepackage[utf8]{inputenc}
\usepackage{float}
\usepackage{mathtools}
\usepackage{tikz}
\usepackage{physics}
\usepackage{natbib}
\usepackage{ragged2e}
\usepackage{xcolor, soul}
\definecolor{Gray}{gray}{0.0}
\sethlcolor{red}
\usepackage{graphicx}
\usepackage{epsfig}
\usepackage{hyperref}
\hypersetup{hidelinks,colorlinks=true,urlcolor=blue,linkcolor=blue,citecolor=blue}
\usepackage{subcaption}
\begin{document}

\title{Tuning for Quantum Speedup in Directed Lackadaisical Quantum Walks}
\author{Pranay Naredi}
\email{pranay7661@gmail.com}
\author{J. Bharathi Kannan}
\email{jbharathi.kannan@students.iiserpune.ac.in}
\affiliation{Indian Institute of Science Education and Research, Dr. Homi Bhabha Road, Pune 411008, India}
\author{M. S. Santhanam}
\email{santh@iiserpune.ac.in}
\affiliation{Indian Institute of Science Education and Research, Dr. Homi Bhabha Road, Pune 411008, India}

\date{\today} 

\begin{abstract}
Quantum walks constitute an important tool for designing quantum algorithms and information processing tasks. In a lackadaisical walk, in addition to the possibility of moving out of a node, the walker can remain on the same node with some probability. This is achieved by introducing self-loops, parameterized by self-loop strength $l$, attached to the nodes such that large $l$ implies a higher likelihood for the walker to be trapped at the node. In this work, {\it directed}, lackadaisical quantum walks is studied. Depending on $l$, two regimes are shown to exist -- one in which classical walker dominates and the other dominated by the quantum walker. In the latter case, we also demonstrate the existence of two distinct scaling regimes with $l$ for quantum walker on a line and on a binary tree. Surprisingly, a significant quantum-induced speedup is realized for large $l$. By tuning the initial state, the extent of this speedup can be manipulated.
\end{abstract}

\maketitle

\section{Introduction} 
\label{sec:Introduction}
Discrete-time quantum walks were formally introduced as quantum analog of classical 
random walks \cite{AhaDavZag1993}, though similar ideas had originated earlier 
as well \cite{Fey1986}.
Apart from being a phenomena of intrinsic interest in physics, in the last three decades, quantum walks have emerged as  an important toolkit for designing novel quantum algorithms. A short list of such problems would include element distinctness \cite{Amb2007}, graph traversal \cite{ChiCleDeo2003}, finding triangles in a network \cite{MagSanSze2007}, spatial search \cite{ChiGol2004,ChaNovAmb2016,Mon2016,DylBenFei2021}, quantum random access memory \cite{RyoKazRyo2021}, nonlinear dynamics \cite{BuaDia20,MenMouLyr20}, community detection on networks \cite{MukHat2020} and centrality measures in multi-layer networks \cite{BotPor2021}. Beyond the quantum algorithms, it was shown that any of the quantum gates could be realized using quantum walks. In 
this sense, quantum walks serve as one of the most powerful primitives for realizing a quantum computer \cite{Childs2009,LovCooEve2010,ChiGosWeb2013,SinChaSar2021,SinAldBal21}. 

The central advantage of quantum walks, in comparison with classical random walks, 
is exemplified by the speedup observed in diffusing through a lattice or 
the time taken to reach a specific node in the case of undirected quantum 
walks \cite{AmbBacAsh2001}. This speedup arises primarily from the quantum interferences exhibited by the walker dynamics.
For instance, for quantum walks on
an infinite lattice, the expectation value of distance travelled after $t$ time steps
is $d_t \sim t$ \cite{AmbBacAsh2001}. This contrasts with the classical random walks for which
$ d_t \sim \sqrt{t}$. Hence, this is an instance of a quadratic speedup over the classical walker. On a ring-type lattice, the metric of interest is the mixing time $t_m$ of the walker, namely, the first time at which the
probability distribution of the walker is nearly uniform over the lattice.
For quantum walks, $t_m \sim t$, while for the corresponding classical walk $t_m \sim t^2$.
In general, such quantum speedup translates into faster solution to a host of
problems that can be posed as search on graphs. Recently, spatial search using continuous-time quantum walks
was shown to be quadratically faster than their classical counterparts \cite{ApeChaSha2022}. For example, the element distinctness problem attempts to 
determine if all the $N$ elements in a given list are distinct. If the decision tree model of 
computation is employed, the time complexity for the best known classical algorithm is 
$O(N \log N)$. In contrast, an algorithm based on quantum walks and search for a marked
node on a graph needs only $O(N^{2/3})$ queries \cite{Amb2007}, and is consistent with the 
known lower bounds obtained for this problem \cite{AarShi2004}. Matrix product verification
is another problem of interest for which a quantum walk based algorithm sub-linear in time
is known \cite{BuhSpa2006}. The general expectation is that the observed speedups in quantum walks can be exploited
to design faster quantum algorithms.

Over the last two decades, many experiments have realized quantum walks in test beds
based on NMR quantum computer \cite{DuLiXu2003}, trapped atoms and ions \cite{KarForCho2009,SchMatSch2009}, and photonics \cite{BroFedLan2010, SchCasPot2010, QiaWanXue2021, EhrKeiMac2021,EspBarMar2022,TanDiShi18}. See Refs. \cite{NevPue2018,GraHeiLeb2016} for recent reviews of photonics based experimental approaches.
Recently, parity-time symmetric quantum walks have been experimentally
realized as a means to implement directed quantum walks on graphs necessary for 
the quantum version of page rank algorithms \cite{WuIzaLi2020}. Spatial search on star graphs
has also been realized in photonic quantum hardware \cite{QuMarWan2022,WanShiXia2020}. Superconducting qubit
based quantum processors have been used to simulate quantum walks and is shown to be strongly affected by
noise \cite{KonCliPao2021}. Recent work has shown that quantum advantage with random walks can be predicted
by machine learning algorithms \cite{MelFedAlo2019}.

For the most part, research was focussed on utilizing quantum speedup accrued from {\it undirected}, continuous or discrete time quantum walks taking place on a line, cycle graph or hypercube. In 2015, the lackadaisical quantum walk was introduced as a discrete analogue of
the classical random walk with self-loops at every node \cite{Wong2015}. The addition of self-loops allows the 
walker to stay on a node with some probability. By construction, this is a lazy walker and is similar to that of a three-state quantum walk \cite{Norio_056112, FalBuaDia21}. Surprisingly, the quantum version of the lackadaisical walker improves the performance of quantum walk algorithms  (at least in some cases) compared to those without self-loops \cite{WanWuXu2017}. This was explicitly demonstrated in the case of a search on a two-dimensional
grid \cite{Wong2018}. As an extension of this idea, quantum walkers can also search multiple marked nodes on a grid with optimal choice of self-loop weights \cite{DesDecFer2021}.
A recent review of results related to lackadaisical quantum walk can be found in Ref. \cite{Chi2020}.
In many applications, directed walks (discrete or continuous time variety) are of particular interest 
as in the case of boolean satisfiability problems \cite{Sch1999,CamVenSal2021},
recommender systems, link prediction, computer vision problems \cite{XiaLiuNie2020},
and centrality measures for quantum hub \cite{BoiGre2021}. 
Hence, it is necessary to study {\it directed}, as opposed to undirected, lackadaisical quantum walks, 
especially to understand the effects of self-loops and whether the speedup reported earlier in a 2D grid is
sustained on other network topologies.

A simple template for standard quantum walks on a directed line shows speedup in comparison to classical walks \cite{HoyMey2009}. The quantum speedup in a lackadaisical quantum walk (LQW) is crucially dependent on the number of self-loops. More generally, for LQW, the question is about the limitations of quantum speedup compared to its classical counterpart and how this speedup is affected by self-loops and the choice of initial states. In this work, we study the directed lackadaisical walks, especially to probe the limits of quantum enhanced speedup. As shown in the rest of this paper, we uncover the existence of two scaling regimes in the dynamics of the lackadaisical quantum walker with respect to self-loop strength. This is valid for dynamics on a directed line and directed binary tree topologies. Further, it is also shown that by varying self-loop strength and a parameter in the initial state, a variety of outcomes can be realized -- ranging from regimes in which classical walk is faster to regimes in which quantum speedup is realized. Taken together, this result provides insight into the origin and tunability of quantum enhanced speedup. To proceed further, in Sec. \ref{sec:lqw} the lackadaisical quantum walks is briefly discussed; in Sec. \ref{sec:dlqw} the main analytical and numerical results for a directed lackadaisical walk on a line and binary tree are obtained. In Sec. \ref{sec:mht} mean hitting times are computed, confirming the results obtained in Sec. \ref{sec:dlqw}. Section \ref{sec:conclusions} summarizes the main results.


\section{Lackadaisical Quantum Walks}
\label{sec:lqw}
\indent Consider a graph with $N$ vertices on which quantum walk is to be executed. Let the maximum degree (number of edges) for any vertex be $d$. A walker at a vertex on the graph can move along the directed edge originating
at that vertex. Let $N$ be the size of the Hilbert-space $\mathcal{H}_P$ associated with the position of the walker, and let $d$ be the size of the coin Hilbert space $\mathcal{H}_C$. Each vertex and its edges on the graph are represented, respectively, in terms of basis states in $\mathcal{H}_P$ and $\mathcal{H}_C$. In each vertex whose degree is less than $d$, self-loops (edges connecting the vertex to itself) can be added such that the degree of every vertex is $d$. To describe a lackadaisical quantum walk, \cite{Wong_2015, Nahimovs_2021}, the dimension of $\mathcal{H}_C$ is augmented by the addition of a self-loop with weight $l$ at each vertex so that $\mathcal{H}_C$ is now a $d + 1$-dimensional Hilbert space. The resulting Hilbert space of LQW is $\mathcal{H}=\mathcal{H}_C\otimes\mathcal{H}_P = \mathbb{C}^{d+1} \otimes \mathbb{C}^N$.

\indent Given this Hilbert-space, a walk operator $\widehat{U}$ can be constructed by defining a coin flip operator $\widehat{C}$ that performs a rotation in ``coin-space'', and a shift operator $\widehat{S}$ that evolves the walker in the position space. The walk operator has the general form
\begin{equation}
\widehat{U} = \widehat{S}.(\widehat{C}\otimes I_N),
\label{eq:Uop}
\end{equation}
in which $\widehat{C}$ is chosen to be a Grover coin and is given by
\begin{equation} \label{eq:coin}
\widehat{C}=2\ket{s_c}\bra{s_c}-I_{d+1},
\end{equation}
where $I_{d+1}$ is the $(d+1)-$dimensional identity matrix and
\begin{equation} \label{eq:coin_state}
\ket{s_c}= \frac{1}{\sqrt{d+l}}\left(\sum_{i=0}^{d-1}\ket{i}+\sqrt{l}\ket{\circlearrowleft}\right),
\end{equation}
the coin subspace is spanned by the basis states 
\begin{equation}
\{\ket{i}:i=0,1,\dots d-1\} ~~ \text{and} ~~ \ket{\circlearrowleft},
\label{eq:coin_basis}
\end{equation} 
in which $\ket{\circlearrowleft}$ represents a self-loop. It is to be noted that the coin state $\ket{s_c}$ is an eigenstate of the coin operator with eigenvalue 1. The shift operator depends 
on the topology of the graph on which the walk is executed.

\indent The walker starts from the vertex denoted by the basis state $\ket{0}$ in the position space, and the initial state for the quantum walk is $\ket{\Psi(0)}=\ket{s_{\alpha}}\otimes\ket{0}$, where $\ket{s_{\alpha}}$ is the initial coin state parameterized by $\alpha$ and is taken to be
\begin{equation} 
\ket{s_{\alpha}}= \frac{1}{\sqrt{d+\alpha}}\left(\sum_{i=0}^{d-1}\ket{i}+\sqrt{\alpha}\ket{\circlearrowleft}\right).
\label{eq:s_alpha}
\end{equation}
To understand the significance of $\alpha$, consider the case of $d=1$. If $\alpha=0$, the initial coin state is 
$\ket{s_0} = \ket{0}$ and it favours forward movement of the walker. However, if $\alpha \gg 1$, then $\ket{s_\alpha} \approx \ket{\circlearrowleft}$ and it strongly favours the trapping of the walker. Thus, $\alpha$ is a parameter that allows tuning of the initial state for an entire range of possibilities from moving forward to trapping the walker. Further, the state $\ket{s_{\alpha}}$ with $\alpha\ne l$ is no more the eigenstate of the coin operator. 
The state of the walker at time $t>0$ is
\begin{equation} \label{eq:state_evolution}
    \ket{\Psi(t)}=\widehat{U}^t ~ \left( \ket{s_{\alpha}}\otimes\ket{0} \right)
\end{equation}
In the position space, the reduced density matrix of the walker can be obtained as
\begin{equation} \label{eq:density_mat}
    \rho_W(t)=\mbox{Tr}_{\rm C}(\ket{\Psi(t)}\bra{\Psi(t)}),
\end{equation}
where $\mbox{Tr}_{\rm C}$ denotes tracing over the coin degree of freedom.
Then, the probability of finding the walker at any vertex $\ket{n} \in \mathcal{H_P}$ will be
\begin{equation} 
    P_n(t)=\bra{n}{\rho_W(t)}\ket{n} = \bra{n}\left({\sum_{i=0}^{d}\bra{i}\ket{\Psi(t)}\bra{\Psi(t)}\ket{i}}\right)\ket{n}
\label{eq:prob_dist_1}
\end{equation}
where the summation is performed over the basis states of the $d+1-$dimensional coin subspace.
In the rest of the paper, quantum walks on different types of directed graphs -- walk on line and binary tree-- will be studied. To begin with, we review the results for a lackadaisical quantum walk on an {\it undirected} line.

\subsection{LQW on an undirected line}
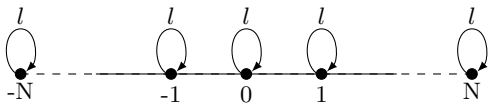
\begin{figure}[b]
\centering

\begin{tikzpicture}
\draw (1,0)--(5,0) node[above]{}; 
\draw[dashed] (0,0)--(6,0) node[above]{}; 

\foreach \x in {-1,0,1}
\filldraw[black] (\x+3,0) circle (2pt) node[below,yshift=-0.06cm]{\x};
\filldraw[black] (0,0) circle (2pt) node[anchor=north]{-N};
\filldraw[black] (6,0) circle (2pt) node[anchor=north]{N};

\foreach \x in {2,3,4}
\draw [-latex] (\x-.05,0) to [out=150, in=180] (\x,0.6) to [out=0, in=45] (\x+.07,0.03) ;
\draw [-latex] (5.95,0) to [out=150, in=180] (6,0.6) to [out=0, in=45] (6.07,0.03);
\draw [-latex] (-0.05,0) to [out=150, in=180] (0,0.6) to [out=0, in=45] (0.07,0.03);
\foreach \x in {2,3,4}
\draw[] node[above,xshift=\x cm,yshift=0.55cm] {$l$};
\draw[] node[above,xshift=0 cm,yshift=0.55cm] {$l$};
\draw[] node[above,xshift=6 cm,yshift=0.55cm] {$l$};

\end{tikzpicture}
\caption{A schematic of undirected line with a self-loop on each vertex having a self-loop weight $l$.}
\label{fig:walk_line}
\end{figure}

\begin{figure}[t]
  \centering
    \includegraphics[width=0.9\linewidth]{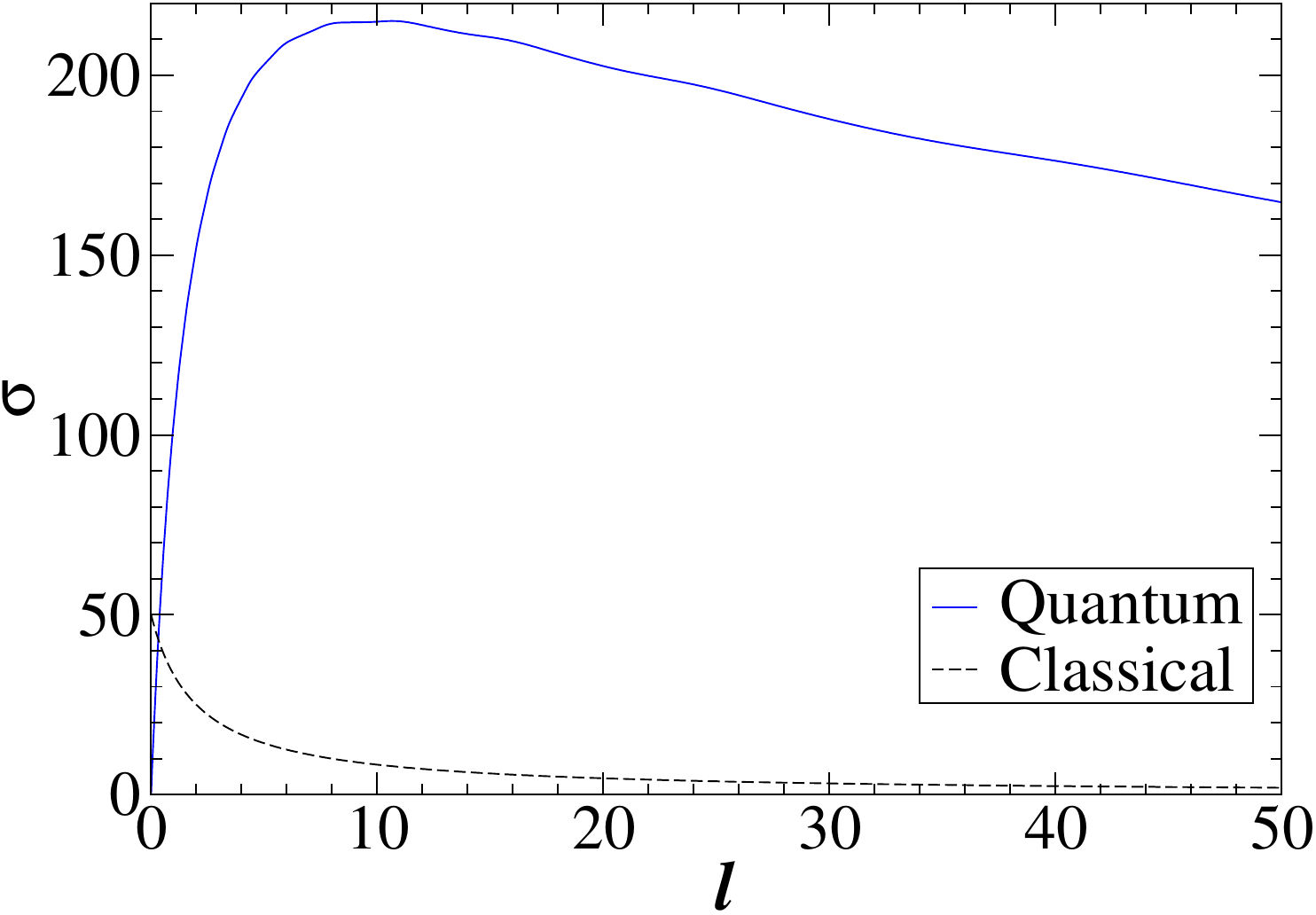}
    \caption{\justifying Spread of the walkers, given by the variance $\sigma^2$ of the position distribution, for the classical and the quantum walker on an undirected line (in $t=50$ iterations) as a function of the self-loop strength $l$. Notice that the quantum walker spreads more, covering a larger distance compared to its classical counterpart, for large $l$.}
    \label{fig:var_uwl}
\end{figure}

\indent Let us first consider a quantum walk on a line \cite{Wang_2017}. The walker starts at the origin denoted by the position ket $\ket{0}$ and can move in both directions with the nearest-neighbour hopping. A schematic of this walk is shown in Fig. \ref{fig:walk_line}. The basis states of $N=2n+1$ dimensional position Hilbert space represents  $N$ vertices $\{0,\pm1,…,\pm n\}$ of the graph. The coin resides in a $3-$dimensional Hilbert space spanned by $\{\ket{\leftarrow},\ket{\rightarrow},\ket{\circlearrowleft}\}$ (here, $\ket{\leftarrow}\equiv \ket{0}$ and $\ket{\rightarrow}\equiv \ket{1}$ from Eq.\ref{eq:coin_state}). The coin has the matrix representation given by
\begin{equation} \label{eq:u1D_coin}
    \widehat{C} = \frac{2}{{2+l}}
                \begin{pmatrix}
                    1&1&\sqrt{l}\\
                    1&1&\sqrt{l}\\
                    \sqrt{l}&\sqrt{l}&l\\
                \end{pmatrix}- 
                \begin{pmatrix}
                    1&0&0\\
                    0&1&0\\
                    0&0&1\\
                \end{pmatrix}
\end{equation}
and the shift operator for the walk is
\begin{equation} \label{eq:line_undirec_shift}
\begin{aligned}
\widehat{S}=\sum_{x=-n}^{n} \ket{\leftarrow}\bra{\leftarrow} &\otimes\ket{x-1}\bra{x} \\
        + \ket{\rightarrow}\bra{\rightarrow} &\otimes\ket{x+1}\bra{x}
        + \ket{\circlearrowleft}\bra{\circlearrowleft} \otimes\ket{x}\bra{x}.
\end{aligned}
\end{equation}
The initial state is chosen to be $\ket{s_{\alpha}}\otimes \ket{0}$. For $l=0$, note that $\widehat{C} = \sigma_x\oplus I_1$ (here Pauli $\sigma_x$ matrix acts as $\sigma_x\ket{\leftarrow}=\ket{\rightarrow}$ and $\sigma_x\ket{\rightarrow}=\ket{\leftarrow}$), so that the evolution of the initial state with any $\alpha$ is an alternating application of $+1$ and $-1$ shift resulting in a trivial walk. If $l>0$, it leads to a non-trivial coin operator.

\indent For comparison purposes, it is instructive to perform a classical random walk on the lattice shown in Fig. \ref{fig:walk_line}. The walker distribution over the lattice can be obtained by evolving the initial state of the classical walker starting from the origin. This is conveniently done using a transfer matrix for $t$ time steps. This probability distribution over the lattice is a Gaussian distribution, and its width $\sigma$ decreases as the self-loop weight $l$ increases (dashed line in Fig. \ref{fig:var_uwl}). This implies that as $l \to \infty$, the distance travelled by the walker diminishes. This behaviour differs from the quantum case in which a large self-loop weight implies that its spread is more compared to the classical walker (blue line in Fig. \ref{fig:var_uwl}).

\section{Directed lackadaisical quantum walk}
\label{sec:dlqw}
\subsection{LQW on directed line}
\begin{figure}[b]
\centering

\begin{tikzpicture}
\draw[dashed] (3,0)--(6,0) node[above]{}; 

\foreach \x in {0,...,3}
\filldraw[black] (\x,0) circle (2pt) node[below,yshift=-0.06cm]{\x};
\filldraw[black] (6,0) circle (2pt) node[anchor=north]{N-1};
\foreach \x in {0,...,3}
\draw [-latex] (\x,0) to [out=0, in=180] (\x+0.91,0);
\draw [-latex] (5,0) to [out=0, in=180] (5.91,0);
\foreach \x in {0,...,3}
\draw [-latex] (\x-.05,0) to [out=150, in=180] (\x,0.6) to [out=0, in=45] (\x+.07,0.03) ;
\draw [-latex] (5.95,0) to [out=150, in=180] (6,0.6) to [out=0, in=45] (6.07,0.03);
\foreach \x in {0,...,3}
\draw[] node[above,xshift=\x cm,yshift=0.55cm] {$l$};
\draw[] node[above,xshift=6 cm,yshift=0.55cm] {$l$};

\end{tikzpicture}
\caption{A schematic of directed walk on a line with a self-loop on each vertex carrying weight $l$}
\label{fig:1D_walk}
\end{figure}
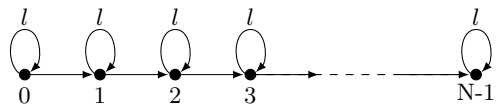

\begin{figure*}
  \centering
\begin{subfigure}{.49\textwidth}
	\includegraphics[width=\linewidth,height=120pt]{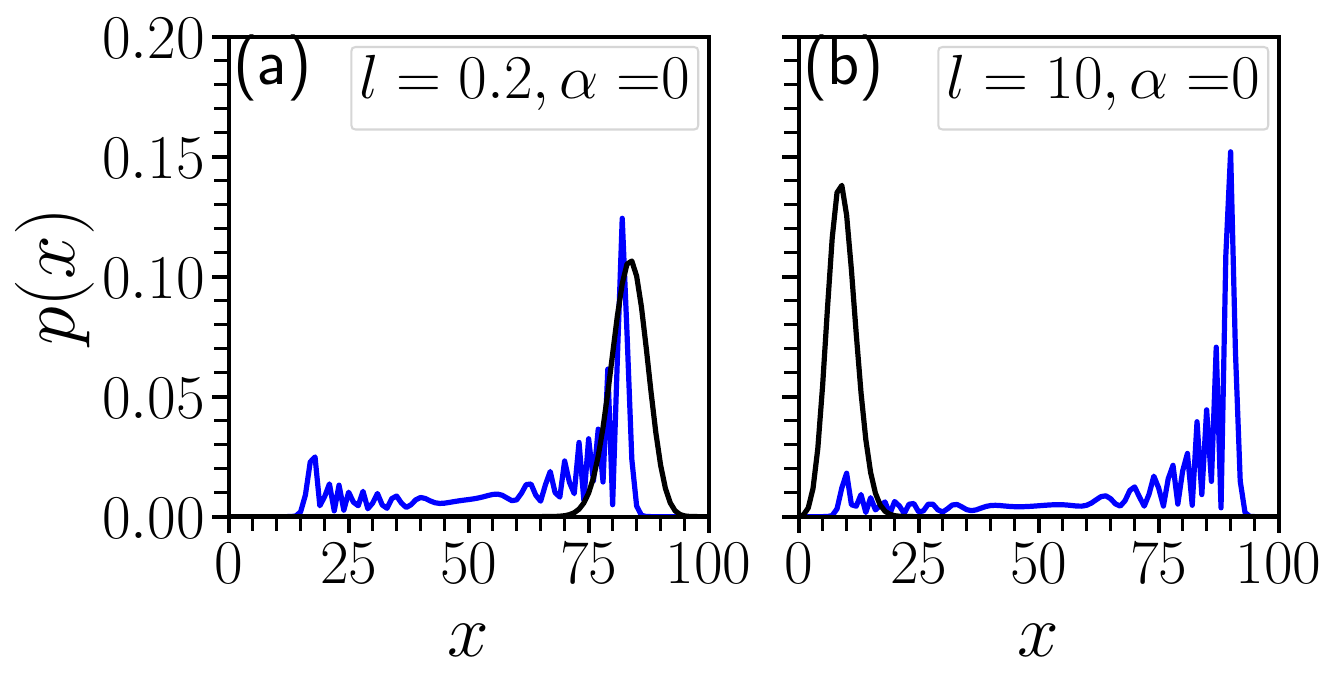}
\end{subfigure}
\begin{subfigure}{0.49\textwidth}
	\includegraphics[width=\linewidth,height=120pt]{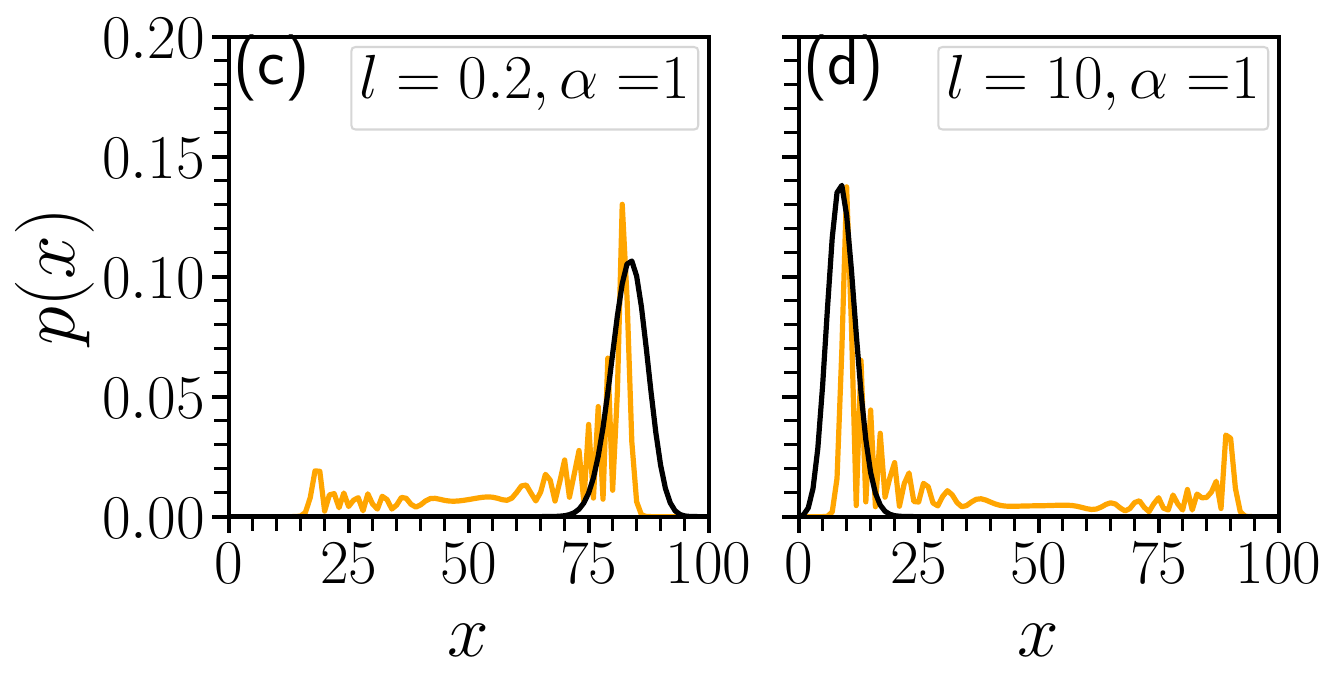}
\end{subfigure}
\begin{subfigure}{0.49\textwidth}
	\includegraphics[width=\linewidth,height=120pt]{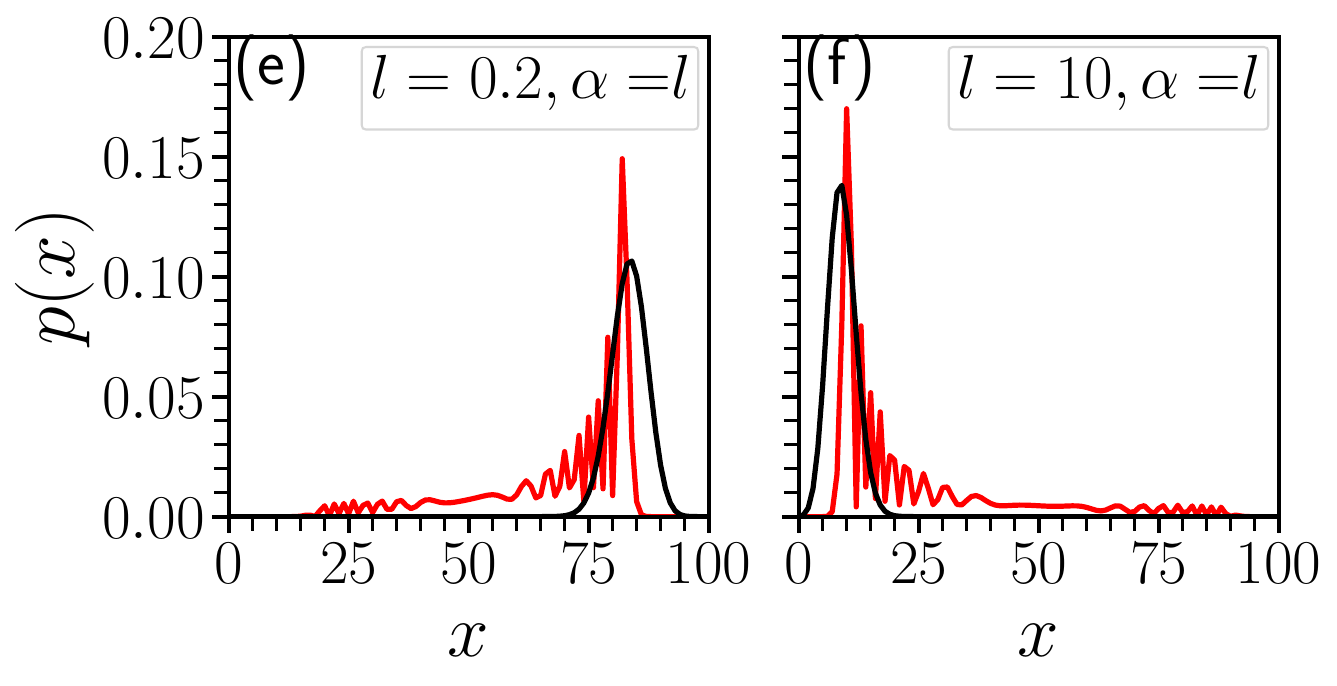}
\end{subfigure}
\begin{subfigure}{0.49\textwidth}
	\includegraphics[width=\linewidth,height=120pt]{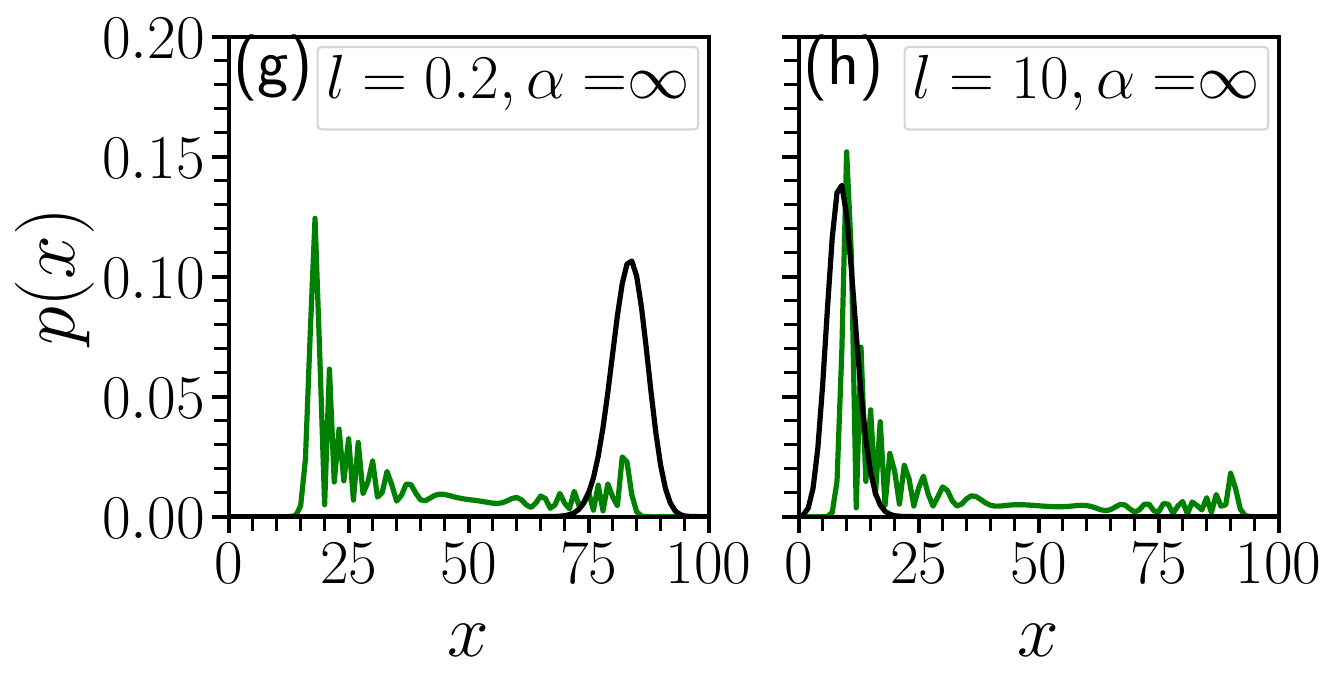}
\end{subfigure}
    \caption{\justifying Comparison of quantum ($\alpha = 0$(blue), $1$(orange), $l$(red), $\infty$ (green)) and the classical (black) walker's probability distribution on a directed-line for $l=10$ (left), $0.2$ (right) at $t=100$.}
    \label{fig:pd_wl}
\end{figure*}

\indent  Next, let us consider a directed quantum walk on a line in which the walker can only advance in one direction, as seen in the schematic in Fig. \ref{fig:1D_walk}. The directed line has $N$ vertices labelled $\{0,1,…, N-1\}$ and represented as the basis states of a $N$-dimensional position Hilbert-space. Periodic boundary conditions are applied for convenience of calculation. The coin subspace $\mathcal{H}_C$ is a $2-$dimensional Hilbert-space spanned by $\{\ket{\rightarrow},\ket{\circlearrowleft}\}$ (here, $\ket{\rightarrow}\equiv \ket{0}$ from Eq.\ref{eq:coin_state}) and has an explicit matrix representation given by
\begin{equation} \label{eq:1D_coin}
    \widehat{C} = \frac{2}{{1+l}}
                \begin{pmatrix}
                    1&\sqrt{l}\\
                    \sqrt{l}&l\\
                \end{pmatrix}- 
                \begin{pmatrix}
                    1&0\\
                    0&1\\
                \end{pmatrix}.
\end{equation}
The shift operator is 
\begin{equation} \label{eq:1D_shift}
\widehat{S} = \sum_{x=0}^{N-1}\ket{\rightarrow}\bra{\rightarrow}\otimes\ket{(x+1) ~{\rm mod} N}\bra{x}+\ket{\circlearrowleft}\bra{\circlearrowleft}\otimes\ket{x}\bra{x}.
\end{equation}
In this form, this is a unitary operator. In our simulations, $N$ is set larger than the number of time steps, such that the boundaries are avoided and hence the effective graph is a line rather than a ring.

If the self-loop weight is $l=0$ or $l=1$, the walk dynamics is trivial because, in the former case, the coin is the two-dimensional Pauli $\sigma_z$ matrix  (where $\sigma_z\ket{\rightarrow}=\ket{\rightarrow}$ and $\sigma_z\ket{\circlearrowleft}=-\ket{\circlearrowleft}$), and in the latter case, it is the Pauli $\sigma_x$ matrix(where $\sigma_x\ket{\rightarrow}=\ket{\circlearrowleft}$ and $\sigma_x\ket{\circlearrowleft}=\ket{\rightarrow}$). Thus, for $l=1$, the evolution proceeds as alternating applications of +1 positional shift and a self-loop. 

Note that the quantum walk constructed in this section is equivalent to a biased walk on an undirected line without self-loops. To see the equivalence, consider an undirected line graph from vertex $-v'$ to $v'$ with the same coin operator, but in the shift operator the self loop is replaced by a left shift. Then, after $v'$ steps the probability of the walker at a position $x$ in the directed case will be equal to the probability at the position $2x-v'$ for undirected walk. Despite this equivalence, the scaling results and the characteristic timescales we demonstrate in this work are entirely new (including for the case of an undirected walker). 

Figure \ref{fig:pd_wl} shows the probability distribution of the walker positions for several choices of self-loop weight $l$. Note that due to the symmetry of the $2-$dimensional coin [Eq. \ref{eq:1D_coin}], the distributions for weights $l$ and $1/l$, after evolution of $t$ time steps are mirror images along the $x=t/2$ line. This is also true for the classical random walk. 

For the undirected quantum walks, the distribution of walker position $p(x)$ is generally bimodal, and hence its spread as measured by its standard deviation $\sigma_x$ gives a good idea about the ``distance'' traveled by the walker from the origin. As $p(x)$ has a spread about the origin, the mean position $\langle x \rangle$ does not provide a good measure of the distance traveled. In contrast, for a directed walker $\langle x \rangle$ is a better indicator of the distance traveled because the corresponding $p(x)$ is not spread about the origin and is usually a sharply-peaked and evolving wavepacket as seen in Fig. \ref{fig:pd_wl}. Therefore, in the rest of this paper, we will use $\langle x \rangle$ as a measure of ``distance travelled'' by the walker from its initial position at $x=0$ on directed graphs.

The classical walker dynamics depend only on $l$ and is straightforward to understand. For $l \gg 1$, the walker is trapped by the self-loops and forward movement is strongly suppressed. If $l < 1$, then the walker moving ahead is favoured. This feature is clearly observed in Fig. \ref{fig:pd_wl} (a-h). Unlike the classical walker, the dynamics of the quantum walker reveal a variety of distinct behaviours depending on the value of $l$ and $\alpha$. To understand this scenario, let us consider the case of LQW with $d=1$ and coin states to be $\{\ket{\rightarrow},\ket{\circlearrowleft}\}$. We recall that the initial state $\ket{\Psi} = \ket{s_\alpha}$ is not an eigenstate of the coin operator $\widehat{C}$, except if $\alpha=l$. Indeed, we have
\begin{equation}
\widehat{C} ~ \ket{s_\alpha} = 2 \ket{s_c} ~ \bra{s_c}\ket{s_\alpha} - \ket{s_\alpha},
\label{eq:coin1}
\end{equation}
where $\bra{s_c}\ket{s_\alpha} = \frac{1+\sqrt{l \alpha}}{\sqrt{(1+l)(1+\alpha)}}.$
For $\alpha=0$ case, using Eqs \ref{eq:coin_state} and \ref{eq:s_alpha} in Eq. \ref{eq:coin1}, we obtain
\begin{equation}
\widehat{C} ~ \ket{s_{\alpha=0}} = \frac{1-l}{1+l} \ket{\rightarrow} + \frac{\sqrt{l}}{1+l} \ket{\circlearrowleft}.
\end{equation}
Now, based on the limiting cases given by
\begin{align}
\widehat{C} ~ \ket{s_{\alpha=0}} & \approx \;\; \ket{\rightarrow}  \;\;\;\; (l \to 0), \nonumber \\
                                 & \approx -\ket{\rightarrow}  \;\;\; (l \gg 1),
\label{eq:coin-a0}
\end{align}
it can be inferred that for $\alpha=0$, the quantum walk does not remain trapped for any value of self-loop weight $l$. This tendency is clearly observed in Fig. \ref{fig:pd_wl}(a,b) for both $l=0.2$ and $l=10$, respectively. Note that the corresponding classical walk is constrained by $l$. A similar scenario unfolds in Fig. \ref{fig:pd_wl}(c,d) for $\alpha=1$ with a minor difference that at $l=10$ the classical and quantum dynamics have not diverged far from one another. This ultimately happens as $l$ increases even more.

If $\alpha=l$, we get the following set of limiting cases
\begin{align}
\widehat{C} ~ \ket{s_{\alpha=l}} & \approx \ket{\rightarrow}  \;\;\; (l \to 0), \nonumber \\
                                 & \approx \ket{\circlearrowleft}  \;\;\;\; (l \gg 1).
\label{eq:coin-al}
\end{align}
This predicts that for $l<1$ quantum walk dynamics would be favoured, but it will be strongly suppressed for $l \gg 1$. This scenario is seen in Fig. \ref{fig:pd_wl}(e,f) for $l=0.2$ and $l=10$, respectively.
For $\alpha \to \infty$, through a similar argument, it can be shown that quantum walk is strongly suppressed since
\begin{align}
\widehat{C} ~ \ket{s_{\infty}} & \approx \ket{\circlearrowleft}  \;\;\; (l \to 0), \nonumber \\
                                 & \approx \ket{\circlearrowleft}  \;\;\; (l \gg 1).
\label{eq:coin-ainf}
\end{align}
This is corroborated by the numerical simulations shown in Fig. \ref{fig:pd_wl}(g,h) for $\alpha=\infty$ with $l=0.2$ and $l=10$, respectively. In numerics, $\alpha=\infty$ is easily implemented by taking the initial coin state to be $\ket{s_{\alpha}}=\ket{\circlearrowleft}$. It is clear that the dynamics of the quantum walker depends on both $\alpha$ and $l$. By varying $\alpha$ and $l$, we can realize two distinct scenarios ; ({\it a}) quantum walker reaches a lattice site faster than its classical counterpart (we will denote this as a quantum speedup), or ({\it b}) the other extreme limit of the classical walker being faster than the quantum walker. Further, in all the cases shown in Fig. \ref{fig:pd_wl}, since the evolving probability distributions remain sharply peaked, the mean position $\langle x \rangle$ is a reasonably good indicator of the distance travelled from $x=0$ and is a convenient metric to track the progression of the quantum walker. Therefore, we will present our results and conclusions based on the evolution of the mean position of the classical and quantum walkers.

\begin{figure}[t]
    \includegraphics[width=\linewidth,height=180pt]{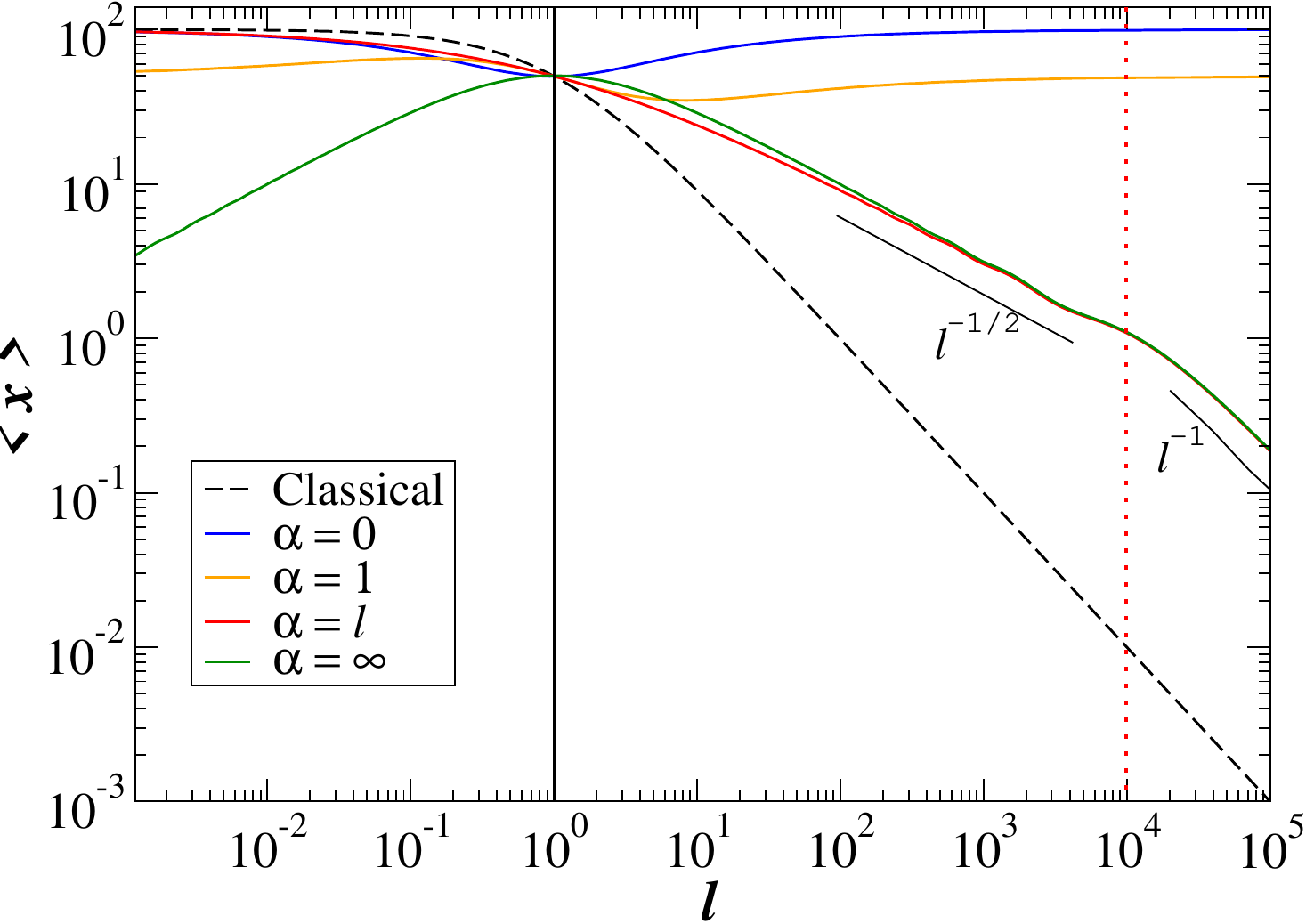}
    \caption{\justifying Mean of the probability distribution of LQW on a directed line as a function of self-loop weight $l$ for $\alpha = 0, 1, l$ and $\infty$ (solid lines). The corresponding classical walk is shown as a dashed (black) line. This result is based on simulations performed for $t=100$ iterations. Two distinct regimes are indicated, namely, $l^{-1/2}$ and $l^{-1}$. The timescale $t^{*}=t/\sqrt{l}=1$ (see Eq. \ref{eq:meanx_tscale}) corresponding to the transition from one regime to the other is indicated by a vertical (dotted, red) line at $l=l^{*}=10^4$. The solid (black) vertical line separates the regime of $l < 1$ and $l > 1$.}
    \label{fig:mean_wl}
\end{figure}

To obtain a broader perspective, Fig. \ref{fig:mean_wl} shows $\langle x \rangle$ at a fixed value of $t=100$ time steps as a function of $l$. For $l <1$, surprisingly classical walker travels farther than the quantum walker for any value of $\alpha$. This is the region to the left of the black vertical line in Fig. \ref{fig:mean_wl}.
For $l=1$, both the classical and quantum walkers cover the same mean distance. Taken together, this regime illustrates the limitations of quantum speedup and shows that quantum walks need not always perform better than the corresponding classical walks. For $l >1$, the classical walk is strongly restricted (region to the right of the black vertical line in Fig. \ref{fig:mean_wl}). In this regime, the quantum speedup is evident as the quantum walk outperforms the classical walk, and the extent of divergence from the classical walk depends on the choice of $\alpha$. Two main features in Fig. \ref{fig:mean_wl} must be pointed out. Based on Eq. \ref{eq:coin-a0}, it can be inferred that for $0 < \alpha \gtrsim 1$, to a first approximation, the mean position of the quantum walk is nearly independent of  $l$, especially as $l \gg 1$. For $\alpha \gg 1$ and $l \gg 1$, we obtain a remarkable result that
\begin{equation}
\langle x \rangle \propto \frac{1}{l}.
\end{equation}
This is identical to the corresponding classical result $\langle x \rangle_{\rm cl} \propto \frac{1}{l}$. Though the mean position of both the classical and quantum walk decays as $l^{-1}$, the quantum walker maintains a mild quantum speedup with respect to the classical walker. Both these features can be seen in the simulation results shown in Fig. \ref{fig:mean_wl}.

Now, we provide analytical proof of $1/l$ decay for the mean position of lackadaisical directed quantum walk defined through the coin operator in Eqs.  \ref{eq:coin}-\ref{eq:coin_basis} with $d=1$, and shift operator in Eq. \ref{eq:1D_shift}. For convenience we use the notations $\ket{0}$ and $\ket{1}$ for $\ket{\rightarrow}$ and $\ket{\circlearrowleft}$, respectively. Note that the coin operator in Eq. \ref{eq:coin} can be written as 
\begin{equation}
\widehat{C} = \dfrac{1-l}{l+1}\sigma_z+\dfrac{2\sqrt{l}}{l+1}\sigma_x.
\end{equation}
After $t$ steps of directed LQW, the state of the system will be $\ket{\Psi(t)} = \widehat{U}^t\ket{\Psi(0)}$, where\\
    \begin{equation} \label{eq:un_line}
    \begin{aligned}
       \widehat{U}^t &= \left[\dfrac{1-l}{l+1} \widehat{S}.\sigma_z\otimes I+\dfrac{2\sqrt{l}}{l+1}\widehat{S}.\sigma_x\otimes I\right]^t.
    \end{aligned}
    \end{equation}
A formal power series expansion of $U^t$ gives
\begin{equation} \label{eq:un_line_N}
    \begin{aligned}
        \widehat{U}^t = \ \left(\dfrac{1-l}{l+1}\right)^t\left(\widehat{S}.\sigma_z\otimes I\right)^t+\left(\dfrac{1-l}{l+1}\right)^{t-1}\left(\dfrac{2\sqrt{l}}{l+1}\right) \\
       \ \sum_{k=0}^{t-1}\left(\widehat{S}.\sigma_z\otimes I\right)^{t-k-1}\left(\widehat{S}.\sigma_x\otimes I\right)\left(\widehat{S}.\sigma_z\otimes I\right)^{k}  + \dots     
       \nonumber
    \end{aligned}
    \end{equation}
In the limit $l \gg 1$, the significant contribution will arise only from the first two terms. To simplify further, we note that 
\begin{align}
& [\widehat{S},\sigma_z\otimes I] = 0, \nonumber \\
& [\widehat{S},\sigma_x\otimes I] = \sum_{x=0}^{N-1}\left(\ket{0}\bra{1}-\ket{1}\bra{0}\right)\otimes\left(\ket{x+1}\bra{x}-\ket{x}\bra{x}\right).
\end{align}

Now, the initial state in \ref{eq:s_alpha} with $d=1$ can be evolved using the evolution operator : 

\begin{equation} \label{eq:st_evol_large_l}
    \ket{\Psi(t)} = \widehat{U}^t\dfrac{1}{\sqrt{\alpha+1}}\left(\ket{0}+\sqrt{\alpha}\ket{1}\right)\otimes\ket{0}
\end{equation}          

 \begin{widetext}
    \begin{align}
        \ket{\Psi(t)} = & \left(\dfrac{l-1}{l+1}\right)^t\dfrac{1}{\sqrt{\alpha+1}}\left((-1)^t\ket{0}\otimes\ket{t} +\sqrt{\alpha}\ket{1}\otimes\ket{0}\right) \nonumber \\
        +&\left(\dfrac{l-1}{l+1}\right)^{t-1}\left(\dfrac{2\sqrt{l}}{l+1}\right)\dfrac{1}{\sqrt{\alpha+1}}\sum_{k=0}^{t-1}(-1)^k\left((-1)^{t-1}\sqrt{\alpha}\ket{0}\otimes\ket{t-k}+\ket{1}\otimes\ket{k}\right)+ \dots
        \label{eq:st_evol_large_l_approx}
    \end{align} 
To compute the mean position $\expval{x}$ of the walker, the position operator in the walker Hilbert-space acts on a walker state as $\widehat{x} \ket{n} = n\ket{n}$, which leads to $\expval{x (t)} = \expval{x} = \bra{\Psi(t)}x\ket{\Psi(t)}$.

\begin{align}
x\ket{\Psi(t)} = & \left(\dfrac{l-1}{l+1}\right)^t\dfrac{1}{\sqrt{\alpha+1}}(-1)^t t\ket{0}\otimes\ket{t} \nonumber \\
+ & \left(\dfrac{l-1}{l+1}\right)^{t-1}\left(\dfrac{2\sqrt{l}}{l+1}\right)\dfrac{1}{\sqrt{\alpha+1}}\sum_{k=0}^{t-1}(-1)^k\left((-1)^{t-1}\sqrt{\alpha}(t-k)\ket{0}\otimes\ket{t-k}+k\ket{1}\otimes\ket{k}\right) + \dots
\end{align}

Using this expression, after some simple manipulations, the mean position can be obtained as
\begin{align}
\bra{\Psi(t)}x\ket{\Psi(t)} =
    		& \left( \dfrac{l-1}{l+1} \right)^{2t}\dfrac{t}{\alpha+1}
    		- \left( \dfrac{l-1}{l+1} \right)^{2t-1} \dfrac{4t\sqrt{l \alpha}}{(l+1)(\alpha+1)}
    		+ \left( \dfrac{l-1}{l+1}\right)^{2t-2} \left( \dfrac{2\sqrt{l}}{l+1} \right)^2 \dfrac{\alpha t^2+\alpha t+t^2-t}{2(\alpha+1)} + \dots \nonumber \\	
    	\overset{l \gg 1}\approx & ~ \dfrac{1}{\alpha+1} \left( t-\dfrac{4\sqrt{\alpha}t}{\sqrt{l}} + \dfrac{1}{l} \left[ \alpha(2t^2+2t)+2t^2-4t \right] \right) + \mathcal{O}\left(l^{-3/2}\right)
\label{eq:meanx_final}	
\end{align}

\end{widetext}
Equation \ref{eq:meanx_final} is the general but approximate expression for the mean position valid for $l \gg 1$ and depends on parameters $\alpha, l$ and $t$. In the limit of $\alpha \gg 1$, such that $t/\alpha \ll 1$ and $t/l \ll 1$, we obtain
\begin{align}
\expval{x} \sim -\frac{4}{\sqrt{\alpha}} \frac{t}{\sqrt{l}} + 2 \frac{t^2}{l}.
\label{eq:meanx_case1}
\end{align}
Clearly, for large $\alpha$, scaled parameter can be identified as $t^{*} = t/\sqrt{l}$. Thus, first term in Eq. \ref{eq:meanx_case1} dominates for $t^{*} < 1$, and the second term dominates for $t^{*} > 1$.  This implies that in the limit of large $l$, $t^{*}$ defines a relevant timescale for transition between the two regimes, and it can be identified as
\begin{align}
\expval{x} & \propto \frac{1}{\sqrt{l}}, \;\;\;\; (t^{*} \ll 1),  \nonumber \\
           & \propto \frac{1}{l},        \;\;\;\;\; (t^{*} \gg 1).
\label{eq:meanx_tscale}
\end{align}
This result is borne out by the numerical results shown in Fig. \ref{fig:mean_wl} for $\alpha \gg 1$. Both regimes are clearly visible in the figure, and the vertical line indicates the timescale $t^{*}$. 

In the opposite limit of $\alpha \ll 1$ such that $t/\alpha \gg 1$, and $l \ll 1$ such that $t/l \gg 1$, Eq. \ref{eq:meanx_final} simplifies to 
\begin{align}
\expval{x} & \approx t.
\label{eq:meanx_approx1}
\end{align}
In this limit, $\expval{x}$ is independent of $l$.
However, if $\alpha \gg 1$ and $l \ll 1$, then we obtain
\begin{align}
\expval{x} & \approx t \left( \frac{1}{1+\alpha} + \frac{4}{\sqrt{\alpha}} \sqrt{l} \right) \propto \sqrt{l}.
\label{eq:meanx_approx2}
\end{align}
We expect the mean position to have $\sqrt{l}$ dependence in this limit.
The result in Eq. \ref{eq:meanx_approx1} is consistent with the simulations shown in Fig. \ref{fig:mean_wl} for $\alpha = 0$ and $\alpha=1$, where $\expval{x}$ does not show any significant dependence on $l$. For $\alpha \gg 1$, Fig. \ref{fig:mean_wl} also reveals a similar agreement with Eq. \ref{eq:meanx_approx2} showing $\expval{x} \propto \sqrt{l}$.
As we shall show below, it is also remarkable that Eq. \ref{eq:meanx_tscale} holds for LQW performed on a binary tree topology.

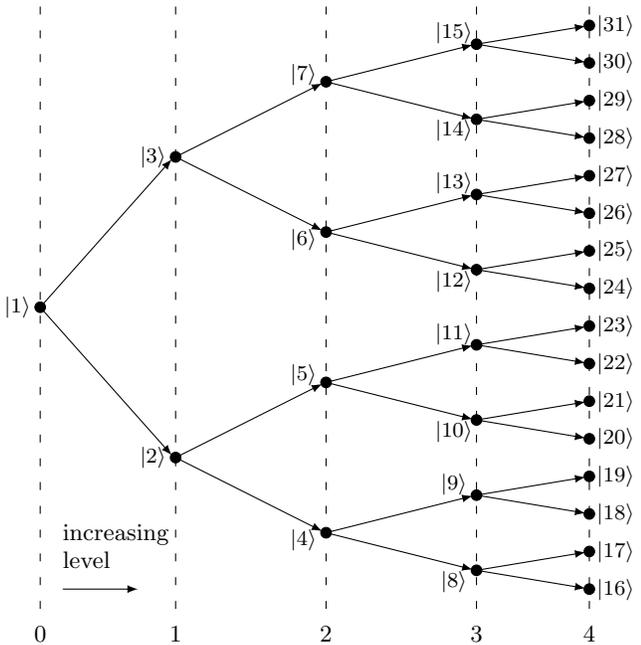
\begin{figure}[t]
\centering
\begin{tikzpicture}

\foreach \x in {0.25,0.75,1.25,...,7.75}{
\filldraw[black] (.5,\x) circle (2pt);
}
\foreach \x in {0.5,1.5,2.5,...,7.5}{
\filldraw[black] (-1,\x) circle (2pt);
}
\foreach \x in {1,3,5,7}{
\filldraw[black] (-3,\x) circle (2pt);
}
\foreach \x in {2,6}{
\filldraw[black] (-5,\x) circle (2pt);
}
\foreach \x in {-6.8}
\filldraw[black] (\x,4) circle (2pt);

\node at (-7.1,4) {\footnotesize $\ket{1}$};
\node at (-5.3,2) {\footnotesize $\ket{2}$};
\node at (-5.3,6) {\footnotesize $\ket{3}$};
\node at (-3.3,1-.1) {\footnotesize $\ket{4}$};
\node at (-3.3,3.1) {\footnotesize $\ket{5}$};
\node at (-3.3,5-.1) {\footnotesize $\ket{6}$};
\node at (-3.3,7.1) {\footnotesize $\ket{7}$};
\foreach \i in {8,10,...,14}
\node at (-1.3,\i-7.5-.15) {\footnotesize $\ket{\i}$};
\foreach \i in {9,11,...,15}
\node at (-1.3,\i-7.5+.15) {\footnotesize $\ket{\i}$};
\foreach \i in {16,17,...,31}
\node at (.85,0.5*\i-7.75) {\footnotesize $\ket{\i}$};

\foreach \i in {1,-1}{
\draw [-latex] (-6.8,4)--(-5-.05,4+1.97*\i);
\foreach \j in {1,-1}{
\draw [-latex] (-5,4+2*\i)--(-3-0.05,4+2*\i+0.98*\j);
\foreach \k in {1,-1}{
\draw [-latex] (-3,4+2*\i+1*\j)--(-1-.05,4+2*\i+1*\j+0.49*\k);
\foreach \l in {1,-1}{
\draw [-latex] (-1,4+2*\i+1*\j+0.5*\k)--(0.45,4+2*\i+1*\j+0.5*\k+0.24*\l);
}}}}

\foreach \i in {1,2,3} {
\draw[loosely dashed] (2*\i-7,8)--(2*\i-7,-0.2) node[below,xshift=0.cm,yshift=0.07cm] {$\i$};
}
\draw[loosely dashed] (-6.8,8)--(-6.8,-0.2) node[below,xshift=0.cm,yshift=0.07cm] {$0$};
\draw[loosely dashed] (.5,8)--(.5,-0.2) node[below,xshift=0.cm,yshift=0.07cm] {$4$};

\node[text width=1cm] (0,0) [below,xshift=-6cm,yshift=1.23cm] {increasing\\level};
\draw [-latex] (-6.5,0.25)--(-5.5,0.25);

\end{tikzpicture}

\caption{Binary tree of depth $d=4$. Total vertices $N=2^{d+1}-1$.}
\label{fig:bt}
\end{figure}

All these results are summarised in Table \ref{meanx_table}, which displays both the $t$ and $l$ dependence of $\expval{x}$. For large self-loop strengths, $l \gg 1$, the walker has a larger tendency to be trapped at a lattice site, and hence we intuitively expect the walker progression to be slow. As this table shows, this is precisely the regime of quantum speedup -- quantum walker is faster than the classical counterpart. Further, depending on $\alpha$, quantum speedup can be tuned with respect to the lackadaisical classical walker.

\begin{table}[b]
\begin{ruledtabular}
\begin{tabular}{|c|c|c|c|c|}
             &  \multicolumn{2}{c|}{$l \ll 1$} &  \multicolumn{2}{c|}{$l \gg 1$} \\
\hline
$\alpha \ll 1$ & $\expval{x} \propto t$ & $\expval{x} \propto l^{0}$ & $\expval{x} \propto t^2$ & $\expval{x} \propto l^{0}$ \\
$\alpha \gg 1$ & $\expval{x} \propto t$ & $\expval{x} \propto \sqrt{l}$ & $\expval{x} \propto t^2$ & $\expval{x} \propto l^{-1}$ \\[1mm]
{\rm Classical} & $\expval{x}_{\rm cl} \propto t$ & $\expval{x}_{\rm cl} \propto l^{0}$ & $\expval{x}_{\rm cl} \propto t$ & $\expval{x}_{\rm cl} \propto l^{-1}$ 
\end{tabular}
\end{ruledtabular}
\caption{\label{tab:ex}\justifying  Summary of the dependence of $\expval{x}$ on $t$ and $l$ in various regimes. For $l \ll 1$, due to pre-factors not shown in this table, classical walks are faster than quantum. The limit of $l \gg 1$ is the regime of quantum speedup in which quantum walks could display mild to quadratic speedup (in comparison to corresponding classical walks) depending on $l$ and $\alpha$.}
\label{meanx_table}
\end{table}

\subsection{LQW on a directed binary tree}
A binary tree is a special type of graph in which each node, starting from a root node, has just two edges, each of which connects to a ``child node''. Figure \ref{fig:bt} shows a schematic of a binary tree of depth $d$, chosen to be 4 in this case. The directed binary tree has $N=2^{d+1}-1$ vertices and they are represented by the states $\{\ket{0},\ket{1}, \dots ,\ket{N-1}\}$ in $\mathcal{H}_P$ as shown in the Fig. \ref{fig:bt}. The Hilbert space $\mathcal{H}_C$ associated with the coin is a $3-$dimensional subspace spanned by $\{\ket{\uparrow},\ket{\downarrow},\ket{\circlearrowleft}\}$. Here, $\ket{\uparrow}$ and $\ket{\downarrow}$ control the shift from a node $\ket{i}$ at a depth $L$ to its child nodes $\ket{2i+1}$ and $\ket{2i+2}$ at depth $L+1$ in the position space. As before, $\ket{\circlearrowleft}$ represents a self-loop. The shift operator, in this case, can be written as
\begin{equation}
\begin{aligned}
    \widehat{S}=\sum_{x=1}^{N} & \ket{\downarrow}\bra{\downarrow} \otimes\ket{(2x+1)~{\rm mod} N}\bra{x} \\
        + & \ket{\uparrow}\bra{\uparrow} \otimes\ket{(2x+2)~{\rm mod} N}\bra{x}\\
        + & \ket{\circlearrowleft}\bra{\circlearrowleft} \otimes\ket{x}\bra{x}.
\end{aligned}
\end{equation}

\begin{figure}[t]
    \includegraphics[width=\linewidth,height=180pt]{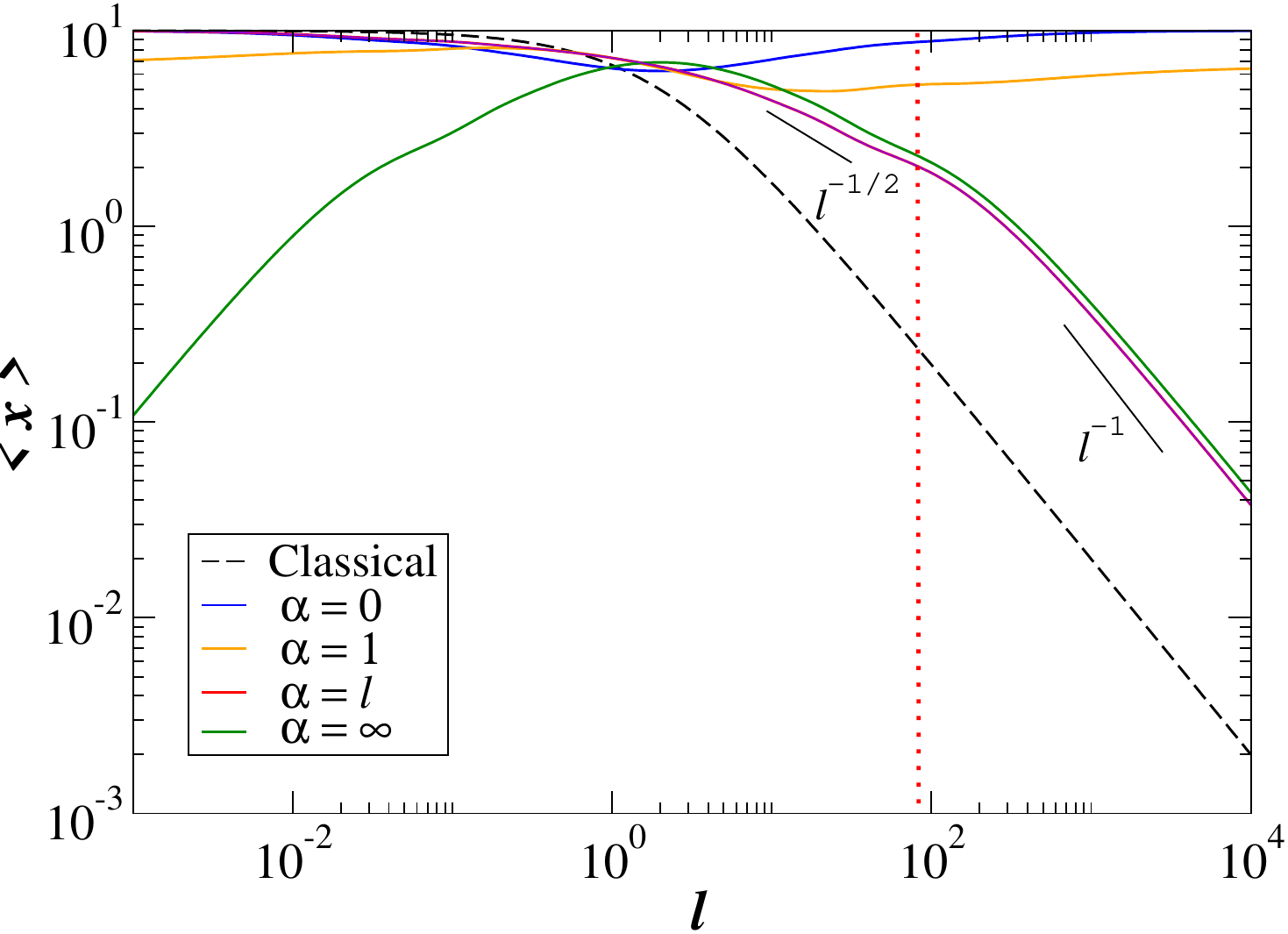}
    \caption{\justifying Mean of the probability distribution of LQW on a binary tree as a function of self-loop weight $l$ for $\alpha = 0, 1, l$ and $\infty$ (solid lines). The corresponding classical walk is shown as a dashed (black) line. This result is based on simulations performed for $t=10$ iterations. Two distinct regimes are indicated, namely, $l^{-1/2}$ and $l^{-1}$. The timescale $t^{*}=t/\sqrt{l}=1$ corresponding to the transition from one regime to the other is indicated by a vertical (dotted, red) line at $l=l^{*}=10^2$.}
    \label{fig:mean_bt}
\end{figure}

The initial state of the quantum walk is $\ket{s_{\alpha}}\otimes\ket{0}$ with $\ket{0}\in \mathcal{H}_P$ being the root node (see Fig. \ref{fig:bt}). The evolved state at time $t$ is given by
\begin{equation} \label{eq:state_evolution_bt}
    \ket{\Psi(t)}=\widehat{U}^t(\ket{s_{\alpha}}\otimes\ket{0}).
\end{equation}
To make the quantum walk directed \cite{10.5555/2011706.2011711}, we can either consider binary trees having infinite depth (i.e., $d\ge t$) or ``terminate'' the walker dynamics once it reaches a leaf node (nodes with no child nodes). This can be achieved with the help of projective measurements at every time step of the evolution. For self-loop weight $l=0$ and $\alpha=0$, the walker reaches the last level ($=d$) with probability one in $t=d$ time steps for both quantum and classical cases. However, for $l > 0$, the probability distribution over position starts to differ from the classical walk, as shown in the Fig. \ref{fig:mean_bt}. In particular, for $l < 1$, the classical walk is efficient, {\it i.e.}, able to traverse more depth than the quantum walk. However, for $l \gg 1$ the quantum walk is more probable to reach the target depth $d$. This is shown in Fig. \ref{fig:mean_bt} for $d=10$ with walks being executed for 10 time steps. In the binary tree, too, as in the case of LQW on a line, we observe a parametric regime in which the quantum walk is less effective than the corresponding classical walk. In general, all the qualitative features of $\expval{x}$ seen in the case of dynamics on a directed line also repeat on the binary tree, for $\alpha < 1$, the mean position is approximately independent of $l$, and this is the regime of quantum speedup. For large $\alpha$ and large $l$, the mean position decays as $l^{-1}$ and yet maintains a mild quantum speedup over the classical walk.
Further, Eq. \ref{eq:meanx_tscale} holds good in this case too. Hence, transition timescale $t^{*} = 1$ yields $l=l^{*} \approx 10^2$ at which $\expval{x}(l)$ makes a transition from $1/\sqrt{l}$ to $1/l$ decay. This timescale is marked by a vertical line in  Fig. \ref{fig:mean_bt} and is consistent with this theoretical estimate. Though many aspects of the directed LQW are qualitatively similar on both a line and a binary tree, the value of $l^{*}$ is the crucial differentiator, and it carries the fingerprint of the topology on which the walk is executed.

\begin{figure}[t]
\centering
\includegraphics*[width=0.5\textwidth]{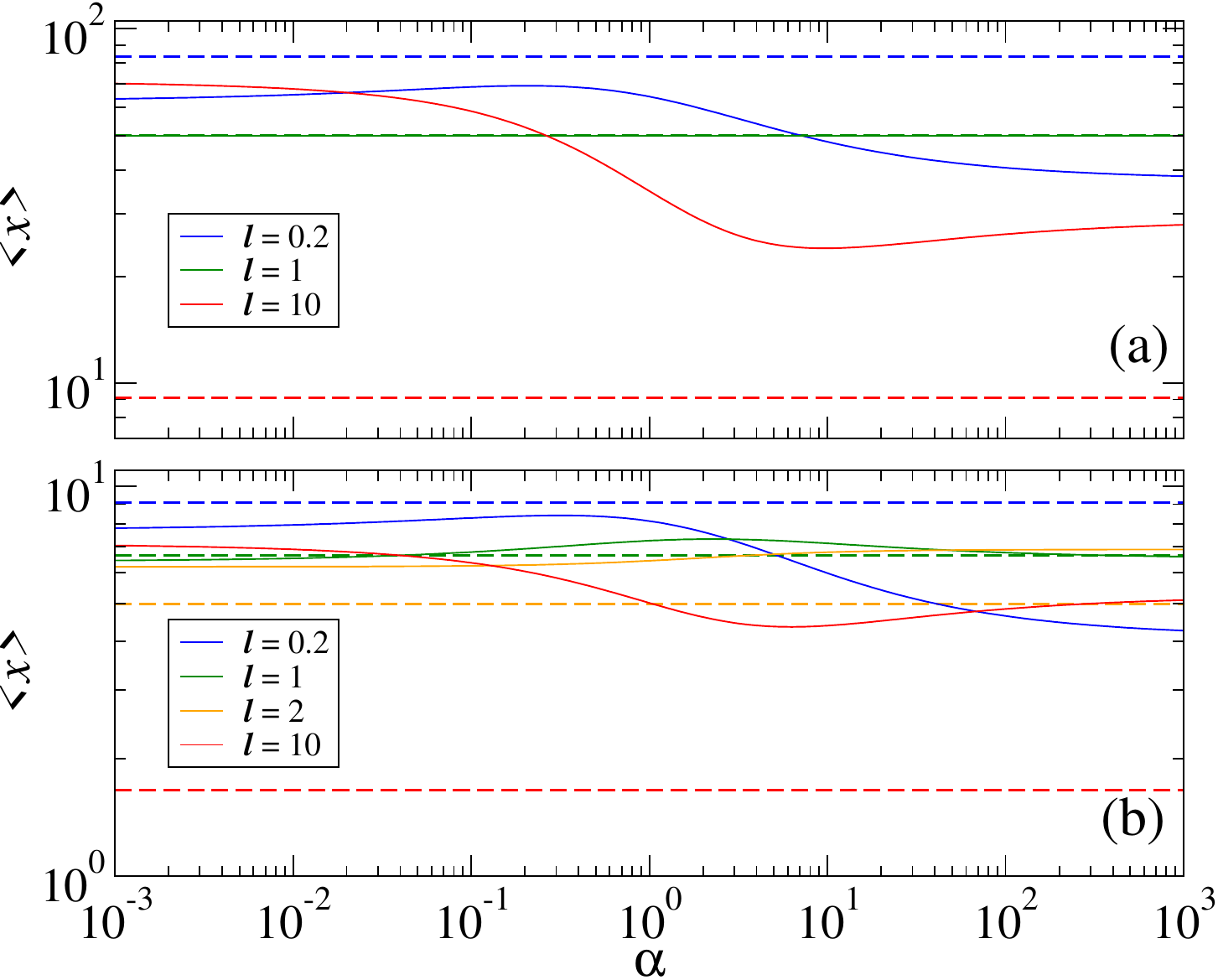}
\caption{\justifying (a) The mean position as a function of $\alpha$ for fixed values of self-loop strength $l$ is shown for walker on a directed line. The solid lines correspond to LQW, and the dashed line corresponds to the classical walk. (b) Same as (a) for a walker on a binary tree. Dashed lines correspond to the classical walker.}
\label{fig:meanx_alpha}
\end{figure}

In all the cases discussed above, it is evident that for directed LQW on the graph with self-loops, there can exist parametric regimes in which quantum speedup is absent, and the classical walk is more efficient. As seen in the simulation results shown in Fig. \ref{fig:meanx_alpha}(a,b), this happens for $l \ll 1$ for any value of $\alpha$. In this limit, the classical walker (shown as a dashed line) is faster than the corresponding quantum walker. At the other extreme, for $l>1$, quantum speedup can be realized to varying degrees depending on $\alpha$ -- from quantum speedup for $\alpha \ll l$ to only a mild speedup for $\alpha \gtrsim l$. This is evident in Fig. \ref{fig:meanx_alpha}(a,b). Thus, by tuning $\alpha$, we are able to tune to a desired level of quantum speedup with respect to the classical walker. A remarkable result is that, for $\alpha, l \gg 1$, the asymptotic mean position of the walker is $\expval{x} \propto l^{-1}$ for all the two topologies reported in this work. In the next section, we employ statistical measures such as the hitting time at a node to emphasize the central result illustrated above.

\section{Mean Hitting time}
\label{sec:mht}
The hitting time is a measure, in an average sense, of the time taken by a walker to reach a particular node.
Using the method developed in \cite{PhysRevA.73.032341} we compute the average hitting time of a quantum walk on a graph having $N$ vertices each with degree $d$. The walker starts from a vertex labelled $x_0$ and stops when it reaches the designated target node $x_f$. Typically, this is performed as follows. A \emph{measured quantum walk} is executed in which the walk operator $\widehat{U}$ and a projective measurement (to check whether the walker has reached $x_f$) are successively applied. Then, the first-crossing probabilities at every time step is determined. The projective measurement has two outcomes, $\widehat{P}$ and $\widehat{Q} = I - \widehat{P}$, where $\widehat{P} =I_d\otimes\ket{x_f}\bra{x_f}$ is the projector onto the final vertex.

\begin{figure}[t]
\includegraphics*[width=0.49\textwidth]{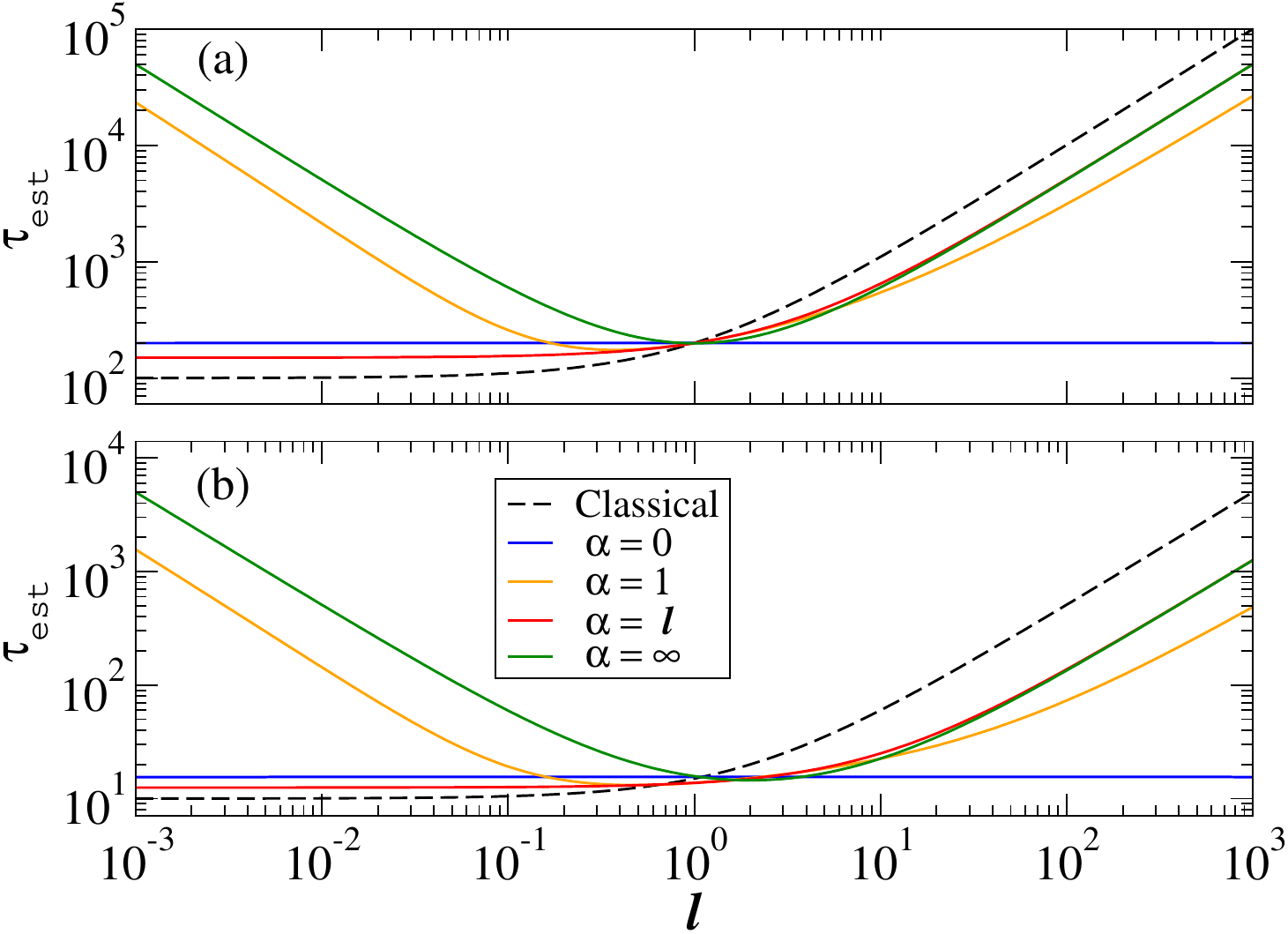}
\caption{\justifying The mean hitting time $\tau_{\rm est}$ as a function of self-loop strength $l$ for lackadaisical walkers on (a) directed line and (b) binary tree. The simulations are shown for several values of $\alpha$ (solid lines). The dashed (black) line is from the corresponding classical walker.}
\label{fig:ht_wl}
\end{figure}
Let us start with an initial density matrix $\rho_0=\ket{\psi_0}\bra{\psi_0}$ in the coin walker space. Then, the first-crossing probability time $t$ would be defined as \cite{PhysRevA.73.032341},
    \begin{equation} \label{eq:cross_p}
    p(t)={\rm Tr}\left( \widehat{P}\widehat{U}[\widehat{Q}\widehat{U}]^{t-1} ~ \rho_0 ~ [\widehat{U^\dagger}\widehat{Q}]^{t-1}\widehat{U^\dagger}\widehat{P} \right).
    \end{equation}
Given $p(t)$, the average hitting time can be obtained as
    \begin{equation} \label{eq:avg_ht}
    \tau = \sum_{t=1}^{\infty}tp(t)
    \end{equation}
In practice, we compute a lower bound for the average hitting time by iterating the quantum walk for the shortest time $T$ such that it satisfies
    \begin{equation} \label{eq:T_sum_p}
    \sum_{t=1}^{T}p(t)\ge 1-\epsilon.
    \end{equation}
Therefore, the mean hitting time can be estimated to be
    \begin{equation} \label{eq:est_ht}
    \tau_{\rm est} = \sum_{t=1}^{T}tp(t).
    \end{equation}\\
As $\epsilon \to 0$, the estimate gets better, {\it i.e.}, $\tau_{\rm est} \to \tau$. 
In the rest of the paper, we use Eq. \ref{eq:est_ht} to compute $\tau_{\rm est}$ LQW on a line and binary tree discussed in Sec \ref{sec:dlqw}. We also compare the quantum hitting time with the classical hitting time, for which the transfer matrix technique is employed for evolving the classical walker. The classical hitting times are calculated similarly to the quantum hitting times.

\emph{Directed walk on a line.} The walker begins from the state $\ket{0}$ and stops when it reaches $\ket{N-1}$. Using Eqs. \ref{eq:Uop} and \ref{eq:1D_coin}-\ref{eq:1D_shift}, we perform a measured quantum walk on the directed line using the projection operators
\begin{equation} \label{eq:1D_proj}
\widehat{P} = I_2\otimes\ket{N-1}\bra{N-1}, \
\;\;\;\;
\widehat{Q} = I_{2N}-\widehat{P}.
\end{equation}
To illustrate the result, we obtain  $\tau_{\rm est}$ for a walk with $N=101$ by summing the series in Eq. \ref{eq:est_ht} with $\epsilon=10^{-6}$. The average hitting times for the quantum and classical walks are displayed in Fig. \ref{fig:ht_wl}(a). For $l < 1$, the classical walk is faster than the quantum walk. As $l$ increases, the quantum walk is faster since the classical walk slows down due to an increasingly higher probability for the walker to use the self-loop. For $l \gg 1$, the mean hitting times for both the classical and quantum walks display a linear relationship with $l$, i.e., $\tau_{\rm est} \propto l$. The hitting time depends on the parameter $\alpha$ of the initial state in a way that is consistent with the results discussed in Sec. \ref{sec:dlqw}.

\emph{Binary tree. } The walker is assumed to start from the root node $\ket{0}\in \mathcal{H}_P$ and stops when it reaches any of the leaf nodes $\{\ket{i}:2^{d}-1\le i \le 2^{d+1}-2\}$. The projection operator for the quantum walk on a perfect binary tree with depth $d$ and $N=2^{d+1}-1$ is
\begin{equation}\label{eq:hyp_proj}
\widehat{P} = I_{3}\otimes\sum_{i=2^d-1}^{2^{d+1}-2}\ket{i}\bra{i}, \;\;\;\; \widehat{Q} = I_{3N}-\widehat{P},
\end{equation}
where $I_{3N}$ is the identity matrix of order $3N$. Figure \ref{fig:ht_wl}(b) shows a comparison of the average hitting times for the classical and quantum walk on the binary tree with depth $d=10$ and $N=2047$ nodes. In this network, too, for $l \ll 1$, the classical walk is faster than the quantum walk, and for $l \gg 1$ quantum walk is faster than the classical. It is also apparent from the figure that the performance of the quantum walker depends on the initial state through the parameter $\alpha$, and hence hitting time can also be tuned by varying $\alpha$. Thus, the hitting times shown in Fig. \ref{fig:ht_wl} are consistent with the analytical and numerical results discussed in Sec. \ref{sec:dlqw}.

\section{Conclusions} 
\label{sec:conclusions}
In this work, we have studied the dynamics of lackadaisical classical and quantum walkers on two directed networks: a line and a binary tree. Lackadaisical quantum walks are similar to the standard quantum walks with an additional self-loop at each node. The self-loop strength $l$ at each node characterizes the probability for the walker to remain on the same node as opposed to transitioning to the neighbouring nodes. In this scenario, intuitively one might expect that as $l$ increases, the walker is more likely to be trapped at some node rather than move ahead. Surprisingly, the work presented here shows that the quantum walker in the large $l$ regime maintains a quantum speedup over the classical walker. The extent of quantum enhanced speedup over the classical walker dynamics - from just about mild to quadratic speedup -- can be tuned by varying the initial state through the parameter $\alpha$ (see Eq. \ref{eq:s_alpha}). For small values of $\alpha$, the quantum walker displays a significantly large quantum speedup. 

It is shown that the lackadaisical quantum walks can be faster or even slower than the corresponding classical walks depending on whether the self-loop weight of the network is larger or smaller than a parameter $l^{*}$. In general, for small $l$, the classical walker is faster compared to its quantum counterpart, and the reverse is true in the large $l$ regime. Based on the analytical and numerical results presented in this paper, it is shown that the distance travelled by the quantum walker exhibits distinct scaling regimes with respect to $l$ (summarised in Table \ref{meanx_table}). For $\alpha \gg 1$, the quantum speedup of a walker executing $t$-steps has two regimes that can be distinguished in terms of the scaled time $t^{*}=t/\sqrt{l}$. For $t^{*} < 1$, the mean position of the walker is proportional to $l^{-1/2}$, whereas for $t^{*} > 1$ it is proportional to $l^{-1}$. Thus, this decay of mean position with $l$ holds good for LQW on a line, binary tree and even quadtree (whose results are not shown here). Even though the mean position decays with $l$, this behaviour can be modified by tuning $\alpha$. For a fixed value of $l$, by varying $\alpha$ it is possible to realize anywhere from mild to quadratic quantum speedup. This effect can be utilized to design better quantum search algorithms on graphs with tunable quantum speedup.



\bibliography{bibliography.bib}


\end{document}